\documentclass[11pt,a4paper]{article}
\usepackage{jheppub}

\title{Probing Sterile Neutrino Parameters with Double Chooz, Daya Bay and RENO}
\author[a]{Kalpana Bora,}
\author[a]{Debajyoti Dutta}
\author[b]{and Pomita Ghoshal}

\affiliation[a]{Physics Department\\
Gauhati University,\\
Assam, India}
\affiliation[b]{Physical Research Laboratory,\\
 Navrangpura, Ahmedabad, India}

\emailAdd{kalpana.bora@gmail.com}
\emailAdd{debajyotidutta1985@gmail.com}
\emailAdd{pomita.ghoshal@gmail.com}

\abstract{In this work, we present a realistic analysis of the potential
of the present-day reactor experiments Double Chooz, Daya Bay and RENO
for probing the existence of sterile neutrinos. We present exclusion
regions for sterile oscillation parameters for each of these experiments, using simulations with realistic estimates of systematic errors and detector resolutions, and compare the 
sterile parameter sensitivity regions we obtain with the existing bounds from other reactor experiments.  We find that these experimental set-ups give significant bounds on the parameter $\Theta_{ee}$ especially in the low sterile oscillation region $0.01 < \Delta m_{41}^2 < 0.05$ eV$^2$. These bounds
can add to our understanding of the sterile neutrino sector since there is still a tension in the allowed
regions from different experiments for sterile parameters.}

\arxivnumber{1206.2172}
\begin{document}
\maketitle

\section{Introduction}

For a long time, it was taken as an accepted fact that there are three
families of fermions, and hence three neutrino flavours. But experiments like LSND \cite{1} and
now MiniBooNE \cite{2} have indicated the presence of a
fourth type of neutrino - the sterile neutrino. Sterile neutrinos
do not carry Standard Model gauge quantum numbers, so they do not take part in Standard Model gauge
interactions, but they can mix with the other 3 active neutrinos.
The LSND result provided evidence for $\bar{\nu}_{\mu}\rightarrow\bar{\nu}_{e}$
oscillations from $\mu^{+}$ decay-at-rest (DAR), and the same oscillation channel was studied in the MiniBooNE
experiment. Similar
results are also indicated by $\bar{\nu_{e}}$ and $\nu_{e}$ disappearance
channels, by the reactor anomaly \cite{3} and Gallium anomaly
\cite{4,5}, and from cosmology \cite{6,7,8,9,10}. These results suggest that the deficit of observed
neutrino fluxes in the respective channels ($\nu_{e}$ or $\bar{\nu_{e}}$)
may be an indication of the existence of a fourth type of neutrino. We
know that the peak energy of $\bar{\nu_{e}}$ produced during beta decay 
in reactors is $\sim$ 3.5 MeV. If the typical baseline of a neutrino
oscillation experiment is $\sim$1 m, then the order of $\Delta m^{2}$
 which could be detected is $\sim\frac{1m}{1MeV}\sim1$ $eV^{2}$.
This is the relevant scale for the oscillation of active neutinos to sterile neutrinos.

The latest global fits of sterile neutrino parameters were presented
in \cite{11,12}. In \cite{11}, 
a fit of all present and past experiments giving bounds on the sterile parameters is performed 
in a (3+1) scenario (i.e., 3 active and 1 sterile neutrino flavors). 
The short baseline neutrino oscillation
results from KARMEN \cite{13,14}, LSND
\cite{15}, $\nu_{e}+{}^{12}C\rightarrow^{12}N_{g.s.}+e^{-}$, as
suggested in \cite{16} have also been included.
Their best fit values at 95 \% CL, including MiniBooNE $\nu_{e}$ and
$\bar{\nu_{e}}$ data, are $\sin^{2}2\Theta_{ee}=0.1$ and $\Delta m_{41}^{2}$=0.9
$eV^{2}$. Earlier, such global analyses on sterile neutrino
parameters in (3+1) or (3+2) scenarios have been presented in \cite{17,18,19,20,21,22,23}. 

A significant amount of work is already available in the literature regarding
the search for sterile neutrinos in reactor and atmospheric neutrino oscillation
experiments. Recently
the possibility of using atmospheric neutrinos as a probe of $eV^{2}$-scale
active-sterile oscillations was explored in \cite{24}, where bounds on $\sin^{2}2\Theta_{\mu\mu}$
and $\Delta m_{41}^{2}$ were presented. 
The implication of sterile neutrinos on measurements of $\theta_{13}$ in a $(3+2)$ scenario in the Double Chooz \cite{25} reactor
experiment was studied in \cite{26}. The impact of light sterile neutrinos
on $\theta_{13}$ measurements in Double Chooz and Daya Bay \cite{27} 
was studied in \cite{28} in a (3+1) scenario. A study of the effect of sterile neutrinos on $\theta_{23}$ and $\theta_{13}$ measurements in MINOS, T2K and Double Chooz 
was performed in \cite{29}. Similar studies 
for Daya Bay were carried out in \cite{30}.
A search for sterile neutrinos using MINOS was
done in \cite{31}. A constraint on the mixing angle $\theta_{14}$ from a combination of Solar and KamLAND data
was given in \cite{32,33}. An analysis of the results of Double Chooz, Daya Bay and RENO to simultaneously fit $\theta_{13}$ and the reactor neutrino anomaly was recently performed in \cite{34}.

In this work, we present exclusion regions in the sterile neutrino parameter
space $\sin^{2}2\Theta_{ee}-\Delta m_{41}^{2}$ for the three
current reactor experiments, namely Double Chooz, RENO \cite{35}
and Daya Bay. A similar study was performed in \cite{28}
for Double Chooz and Daya Bay, where a more approximate analysis 
in the $\sin^{2}2\Theta_{ee}-\Delta m_{41}^{2}$ plane was done assuming an overall
systematic error and neglecting detector resolution. In our present work, we have used simulations with reduced values of errors, as quoted
in the technical reports of the individual reactor experiments, where the cancellation
of correlated reactor-related errors by using both near and far detectors
has been taken into account. Also, we have considered realistic detector resolutions,
which play an important part in the sensitivity analysis.
The allowed regions
in $\sin^{2}2\Theta_{ee}-\Delta m_{41}^{2}$ plane presented
in this work are relevant in view of the fact
that there is still significant tension in the existing data between
appearance and disappearance experiments. Although an exclusion region
is quoted in the present global best-fit scenario \cite{11}, we
have tried to survey critically the region which would be accessible by this
specific set of new reactor experiments, to see what it adds to existing
information. We also compare our results with those of older reactor
experiments like BUGEY \cite{36}, GOSGEN \cite{37} and Krasnoyarsk
\cite{38}.

The paper has been organized as follows. In Section 2, we briefly present
the theory of neutrino oscillations with sterile neutrinos. In Section
3, we describe the details of the three reactor experiments we study, as listed in the
technical reports of each experiment.
We also give details of systematic errors and detector
resolutions used in our analysis. In Section 4, we present
our results, and conclude with a discussion of the results in Section 5.

\section{Theory of Neutrino Oscillations with Sterile Neutrinos}

We know that neutrino mass and flavor eigenstates can be related by
the relation\begin{equation}
\nu_{\alpha}=U_{{\alpha}i}\nu_{i},\end{equation}
 where U is a unitary matrix. In matrix form, this mixing between
flavor eigenstates $\nu_{\alpha}(\alpha=e,\mu,\tau,s$; where s stands
for sterile neutrino) and mass eigenstates $\nu_{j}(j=1,2,3,4)$ in
four neutrino scenario can be represented as \begin{equation}
\begin{pmatrix}\nu_{e}\\
\nu_{\mu}\\
\nu_{\tau}\\
\nu_{s}\end{pmatrix}=\begin{pmatrix}U_{e1} & U_{e2} & U_{e3} & U_{e4}\\
U_{{\mu}1} & U_{{\mu}2} & U_{{\mu}3} & U_{{\mu}4}\\
U_{{\tau}1} & U_{{\tau}2} & U_{{\tau}3} & U_{{\tau}4}\\
U_{s1} & U_{s2} & U_{s3} & U_{s4}\end{pmatrix}\begin{pmatrix}\nu_{1}\\
\nu_{2}\\
\nu_{3}\\
\nu_{4}\end{pmatrix}\end{equation}
 This unitary matrix can be parametrized \cite{39} as \begin{equation}
U=R_{34}(\theta_{34},0)R_{24}(\theta_{24},0)R_{23}(\theta_{23},\delta_{3})R_{14}(\theta_{14},0)R_{13}(\theta_{13},\delta_{2})R_{12}(\theta_{12},\delta_{1})\end{equation}
 where $R_{ij}(\theta_{ij},\delta_{k})$ is the complex rotation matrix
in the i-j plane and the elements are given by \begin{equation}
[R_{ij}]_{pq}=\left\{ \begin{array}{rcl}
\cos\theta\qquad p=q=i & \mbox{or} & p=q=j\\
1\qquad p=q\neq i & \mbox{and} & p=q\neq j\\
\sin\theta e^{-i\delta}\qquad p=i & \mbox{and} & q=j\\
-\sin\theta e^{i\delta}\qquad p=j & \mbox{and} & q=i\\
0\qquad & \mbox{otherwise}\end{array}\right.\end{equation}
 Here, $\theta_{ij}$ is the angle of rotation in i-j plane \cite{17,39}.
If we assume $\theta_{14}, \theta_{24}, \theta_{34} = 0$
then the above mixing matrix reduces to the standard Pontecorvo-Maki-Nakagawa-Sakata
(PMNS) \cite{40,41,42} form. Here
we consider the (3+1) scenario, which is an extension of the three-neutrino scenario
with the addition of one massive sterile neutrino.
Solar and atmospheric neutrino analyses strongly discourage
the (2+2) scenario due to the absence of sterile signals in the atmospheric parameters \cite{43}.
The (3+2) scenario is favoured by the tension between appearance and disappearance experiments in the (3+1) case, but disfavoured by cosmology.

The three new reactor experiments we have considered have two detectors,
one at the far site and the other at the near site. In these
experiments, the electron antineutrino disappearance probability is measured through electron antineutrino events.
The far detector is placed at sufficient distance from the reactor
site (compared to the near one) so as to maximise the disappearance 
of the electron anti-neutrino. Such a two-detector set up has a great advantage
over a single detector, as it can cancel or reduce the systematic
uncertainties. 

In the presence
of a sterile neutrino, the standard 3-flavor neutrino oscillation picture
changes, and hence the survival probability must be modified due to the effect of (3+1) mixing. 
{{The baselines relevant to these experiments are of the order of a few hundred
metres, and hence the oscillation would show signatures at the atmospheric scale as well
as possible sterile-scale effects.}}
The electron neutrino survival probability expression relevant for our analysis
is \cite{28} - 

\begin{equation}
P_{ee}=1-\cos^{4}\theta_{14}\sin^{2}2\theta_{13}\sin^{2}(\frac{\Delta{m}_{13}^{2}L}{4E})-\sin^{2}2\theta_{14}\sin^{2}(\frac{\Delta{m}_{14}^{2}L}{4E}),\end{equation}\\
s

where $L$ is the baseline and $E$ is the neutrino energy. Here,
$P_{ee}$ is seen to be a function of two mass-squared differences
$(\Delta{m}_{14}^{2}$ and $\Delta{m}_{13}^{2})$ and two mixing angles
$\theta_{13}$ and $\theta_{14}$. When
$\theta_{14}\rightarrow0$, we recover the standard three-flavour probability
expression in which the solar mass scale is neglected.

\section{Details of the experiments}

In this section, we present some technical details of the experiments for the
sake of completeness, collected from \cite{25,27,35}.

\subsection{Double Chooz experiment}

The Double Chooz reactor experiment \cite{25,35} is designed to detect $\bar{\nu}_{e}$
through the inverse $\beta$ reaction-\begin{equation}
\bar{\nu}_{e}+p\rightarrow e^{+}+n\end{equation}
 The anti-neutrino flux coming from the two nuclear cores of the Chooz
power plant results from the $\beta$ decay of the fission products of
four main isotopes- $^{235}U$, $^{239}Pu$, $^{241}Pu$ and $^{238}U$.
The Double Chooz far detector is cylindrical in shape, having a radius
of 115 cm and a height of 246 cm, and hence a volume of 10.3 $m^{3}$ . Both
near and far detectors are identical inside the PMT support structure
and the mass of each is about 10.16 tons. The reactor site contains
two reactors, each producing a thermal power of 4.27 GW. The near detector
is situated roughly at a distance of 410 m from the reactor site, while the
far one is at a distance of 1067 m. The reactor core-to-detector distances are
tabulated in table 1.\\

\begin{table}[!t]
\begin{center}
\begin{tabular}{|l|c|r|}
\hline 
Reactor No  & Near detector(km)  & Far detector(km) \tabularnewline
\hline 
1  & 0.47  & 1.12 \tabularnewline
\hline 
2  & 0.35  & 1.00 \tabularnewline
\hline 
\end{tabular}
\caption{Core to detector distances in Double Chooz}
\label{Table 2}

\par\end{center}
\end{table}

In our analysis, we have
considered an exposure time of three years with a 12$\%$ detector resolution.
The $\chi^2$ function we have used is taken from the
collaboration report.

\subsection{Daya Bay}

The Daya Bay neutrino experiment \cite{27} works with reactor generated
electron antineutrinos and uses a gadolinium (Gd) loaded
liquid scintillator detector.
It has two pair of reactors at Daya Bay and Ling Ao I, which generate
11.6 GW of power. One more reactor site at Ling Ao II is under construction.
Daya Bay consists of three underground experimental halls, one far
and two near, linked by horizontal tunnels. Eight identical cylindrical
detectors are employed to measure the neutrino flux. The mass of each detector
is about 20 tons. Four of these eight detectors are at the far zone
while two detectors are kept in each near zone. The distance of the
detectors from the reactor cores at the Daya Bay site is 363 m while this
distance at the Ling Ao site is 481 m. The far detectors are at 1985 m and
1615 m respectively from the Daya Bay and Ling Ao reactor sites. We
have used an exposure time of 3 years, with 12$\%$ resolution.

\subsection{RENO}

The RENO experiment \cite{35} is designed to search for reactor antineutrino disappearance using two identical
detectors. The set-up consists of a near 
detector roughly 292 m away from the reactor array center, while
 the far detector is  about 1.4 km away from the reactor
array center. This design of identical detectors at the two sites
helps in reducing systematic errors. Each detector contains 16 tons
of liquid scintillator which is doped with gadolinium. 
 There are six reactor cores in RENO, for which the core to detector
 distances are listed in Table 2.\\
 
\begin{table}[!t]
\begin{center}
\begin{tabular}{|l|c|r|}
\hline 
Reactor No  & Near detector(km)  & Far detector(km) \tabularnewline
\hline 
1  & 0.70  & 1.52 \tabularnewline
\hline 
2  & 0.48  & 1.43 \tabularnewline
\hline 
3  & 0.32  & 1.39 \tabularnewline
\hline 
4  & 0.32  & 1.39 \tabularnewline
\hline 
5  & 0.48  & 1.43 \tabularnewline
\hline 
6  & 0.70  & 1.52 \tabularnewline
\hline

\end{tabular}
\caption{Core to detector distances in RENO}
\label{Table 2}

\par\end{center}
\end{table}

\begin{table}[!t]
\begin{center}
\begin{tabular}{|l|c|c|r|}
\hline 
Name of Exp  & Double Chooz  & Daya Bay  & RENO \tabularnewline
\hline 
Location  & France  & China  & Korea \tabularnewline
\hline 
No of Reactor cores  & 2  & 4  & 6 \tabularnewline
\hline 
Total Power(GW$_{th}$)  & 8.7  & 11.6  & 16.4 \tabularnewline
\hline 
Baselines- near/Far(m)  & 410/1067  & 363(481)/1985(1615)  & 292/1380 \tabularnewline
\hline 
Target mass(tons)  & 10/10  & 40 $\times$ 2/10  & 16.1/16.1 \tabularnewline
\hline 
No of Detectors  & 2  & 2  & 2 \tabularnewline
\hline 
Exposure(years)  & 3  & 3  & 3 \tabularnewline
\hline 
Resolution($\%$)  & 12  & 12  & 12 \tabularnewline
\hline
\end{tabular}
\caption{Details of the three reactor experiments}
\par\end{center}
\end{table}

The average total thermal power output of the six reactor cores is
16.4 GW$_{th}$, with each reactor core generating about equal power.
The energy of the antineutrinos in these experiments is in the range of
1 to 8 MeV. The resolution is 12$\%$. 

We present the {{particulars of the three reactor experiments in Table 3}}.\\

{{The systematic errors associated with each experimental set-up
are listed in Table 4.{\footnote{Since we did not find specific values of the scaling and overall normalization errors of the Daya Bay experiment in the literature, we have assumed values similar to Double Chooz.}}\\

\begin{table}[!h]
\begin{center}
\begin{tabular}{|l|l|c|c|}
\hline 
Name of Exp  & Double Chooz  & Daya Bay  & RENO \tabularnewline
\hline 
Reactor correlated error($\%$)  & 2.0  & 2.0  & 2.0 \tabularnewline
\hline 
Detector normalisation error($\%$)  & 0.6  & 0.5  & 0.5 \tabularnewline
\hline 
Scaling or calibration error($\%$)  & 0.5  & 0.5  & 0.1 \tabularnewline
\hline 
Overall normalization error($\%$)  & 2.5  & 2.5  & 2.0 \tabularnewline
\hline 
Isotopic abundance error($\%$)  & 2.0  & 2.0  & 2.0 \tabularnewline
\hline
\end{tabular}
\caption{Systematic errors associated with the three experiments}
\par\end{center}
\end{table}

\section{Details of old experiments}

{{The details of the three old reactor experiments ( BUGEY \cite{36},
Gosgen \cite{37} and Krasnoyarsk \cite{38}) with which we compare
our exclusion regions for sterile parameters are tabulated in Table 5.}}\\

\begin{table}[!t]
\begin{center}
\begin{tabular}{|l|l|c|c|}
\hline 
Name of Exp  & Bugey  & Gosgen  & Krasnoyarsk \tabularnewline
\hline 
No of Reactor cores  & 4  & 1  & 3 \tabularnewline
\hline 
Total Power(GW$_{th}$)  & 2.8  & 2.8  & 2.8 \tabularnewline
\hline 
Baselines(m)  & 15,40,95  & 37.9,45.9,64.7  & 57,57.6,231.4 \tabularnewline
\hline 
Target mass($\approx$tons)  & 1.67  & 0.32  & 0.4 \tabularnewline
\hline 
No of Detectors  & 1  & 1  & 1 \tabularnewline
\hline 
Exposure($\approx$ years)  & 0.2  & 0.39,0.56,0.98  & 0.06 \tabularnewline
\hline 
Resolution($\%$)  & 6  & -  & - \tabularnewline
\hline
\end{tabular}
\caption{ Details of old reactor experiments }
\par\end{center}
\end{table}

\section{Results}

We have generated the 90$\%$ c.l. exclusion plots for sterile oscillations for
all the three current experiments and compared the results with the old reactor
experiments Bugey, Gosgen and Krasnoyarsk. The right side of
each contour shows the no-oscillation region while the region left to
the contours is the possible oscillation region. The results are found to be very sensitive
to the values of systematic errors. In our calculations, we have used GLoBES \cite{44,45,46,47,48} for simulating the experiments. 
{{The details of the statistical analysis are as follows: the total no of bins used in all three experiments
 are 62 and the width of the energy window is 1.8-8 MeV.
 The resolution used is 12 $\%$. The uncertainty associated with the shape of neutrino energy spectrum for
 all the three experiments is 2 \%. In our calculation we have used the $\chi^2$ function as defined in the GLoBES manual
and used in standard GLoBES sensitivity analysis:
\begin{equation}
\chi^{2}=\underset{i=1}{\sum^{\#bins}}\underset{d=N,F}{\sum}\frac{(O_{d,i}-(1+a_{R}+a_{d})T_{d,i})^{2}}{O_{d,i}}+\frac{a_{R}^{2}}{\sigma_{R}^{2}}+\frac{a_{N}^{2}}{\sigma_{N}^{2}}+\frac{a_{F}^{2}}{\sigma_{F}^{2}}
\end{equation}
 where $O_{N,i}, O_{F,i}$ are the event rates for the $i^{th}$ bin in the near and far detectors, calculated for
 true values of oscillation parameters; $T_{d,i}$ are the expected event rates for the $i^{th}$ bin in the near and far detector for
 the test parameter values; $a_R, a_F, a_N$ are the uncertainties associated with the reactor flux and detector mass; and 
$\sigma_R, \sigma_F, \sigma_N$ are the respective associated standard deviations.}} 

\begin{figure}[!h]
\begin{centering}
\begin{tabular}{cc}
\includegraphics[width=4.5in]{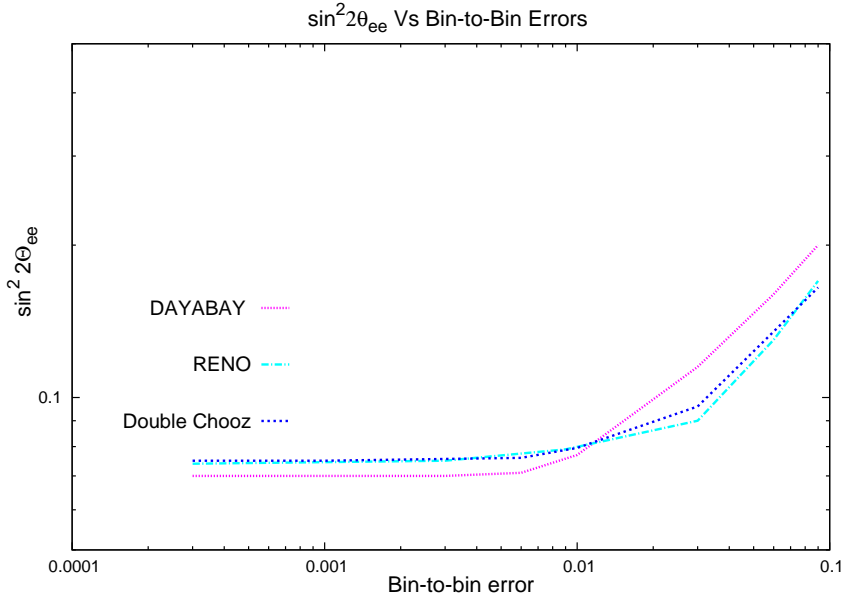} & \tabularnewline\\
\end{tabular}
\par\end{centering}

\caption{ Dependence of the average $\sin^{2}2\theta_{ee}$ bound on the correlated bin-to-bin error for Double Chooz, Daya Bay and RENO.
}

\end{figure}

{{Fig.1 shows the variation of the average bound on $\sin^{2}2\theta_{ee}$ from each of the three experiments Double Chooz, Daya Bay and RENO as a function of the 
bin-to-bin systematic error for a constant overall normalisation. We depict the behavior with respect to the correlated bin-to-bin error since this is found to have the most significant effect on parameter sensitivities. 
The figure shows that the dependence of the sensitivity in the case of Daya Bay is slightly steeper, even for low values of the bin-to-bin systematics, than for the other two experiments. 
}} \\

\begin{table}[!t]
\begin{center}
\begin{tabular}{|l|l|c|c|}
\hline 
Name of Exp  & Double Chooz  & Day Bay  & RENO \tabularnewline
\hline 
Reactor correlated error($\%$)  & 0.06  & 0.087  & 0.5 \tabularnewline
\hline 
Detector normalisation error($\%$)  & 0.06  & 0.12  & 0.5 \tabularnewline
\hline 
Scaling or calibration error($\%$)  & 0.5  & 0.5  & 0.1 \tabularnewline
\hline 
Overall normalization error($\%$)  & 0.5  & 0.5  & 0.5 \tabularnewline
\hline 
Isotopic abundance error($\%$)  & 0.06  & 0.087  & 0.5 \tabularnewline
\hline
\end{tabular}
\caption{Set of reduced errors used in our calculation}
\par\end{center}
\end{table}
\begin{figure}[!h]
\begin{centering}
\begin{tabular}{c}
\includegraphics[width=4.5in]{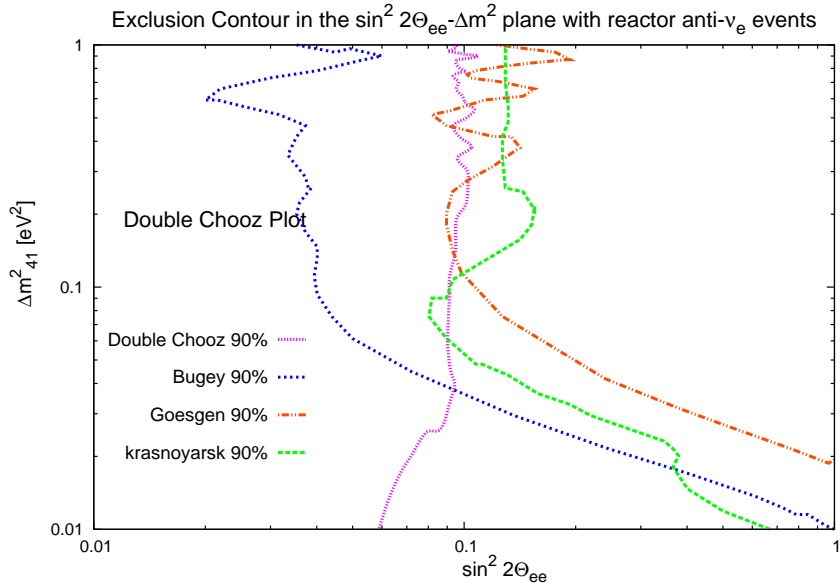} \tabularnewline
\end{tabular}
\par\end{centering}

\caption{Comparison of Double Chooz exclusion plot with other reactor experiments
at 12$\%$ resolution using modified errors. }

\end{figure}

In our further analysis, we have used a reduced set of systematic
errors as listed in Table 6, {{taking into account the partial cancellation of errors due to the presence of 
both near and far detectors, as documented in the experiment literature.}}\\

{{We have included the changed flux normalizations given by the reactor flux anomaly \cite{3}
in our sensitivity analysis. We are performing a simulation and not a data analysis, and hence varying the flux
normalizations has minimal effect on our results, since the relative normalizations simultaneously affect
both the true and test spectra in the simulation and lead to a cancellation of their effect in the
parameter sensitivity. For the same reason, leaving the flux normalization as a freely varying parameter does not have a major
influence on our results, and therefore they may be taken to be indicative of the effect of spectral information only.}}   
\begin{figure}[!h]

\begin{centering}
\begin{tabular}{cc}
\includegraphics[width=4.5in]{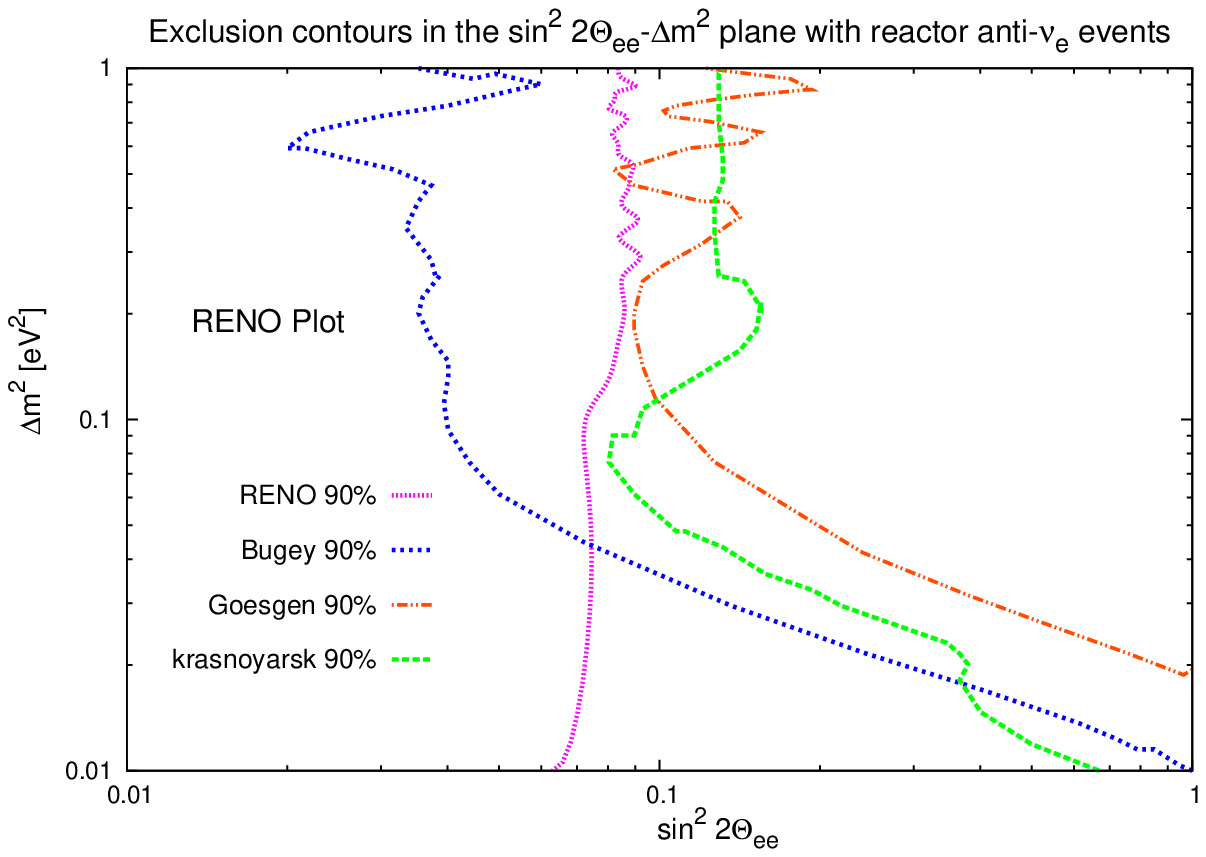} & \tabularnewline
\includegraphics[width=4.5in]{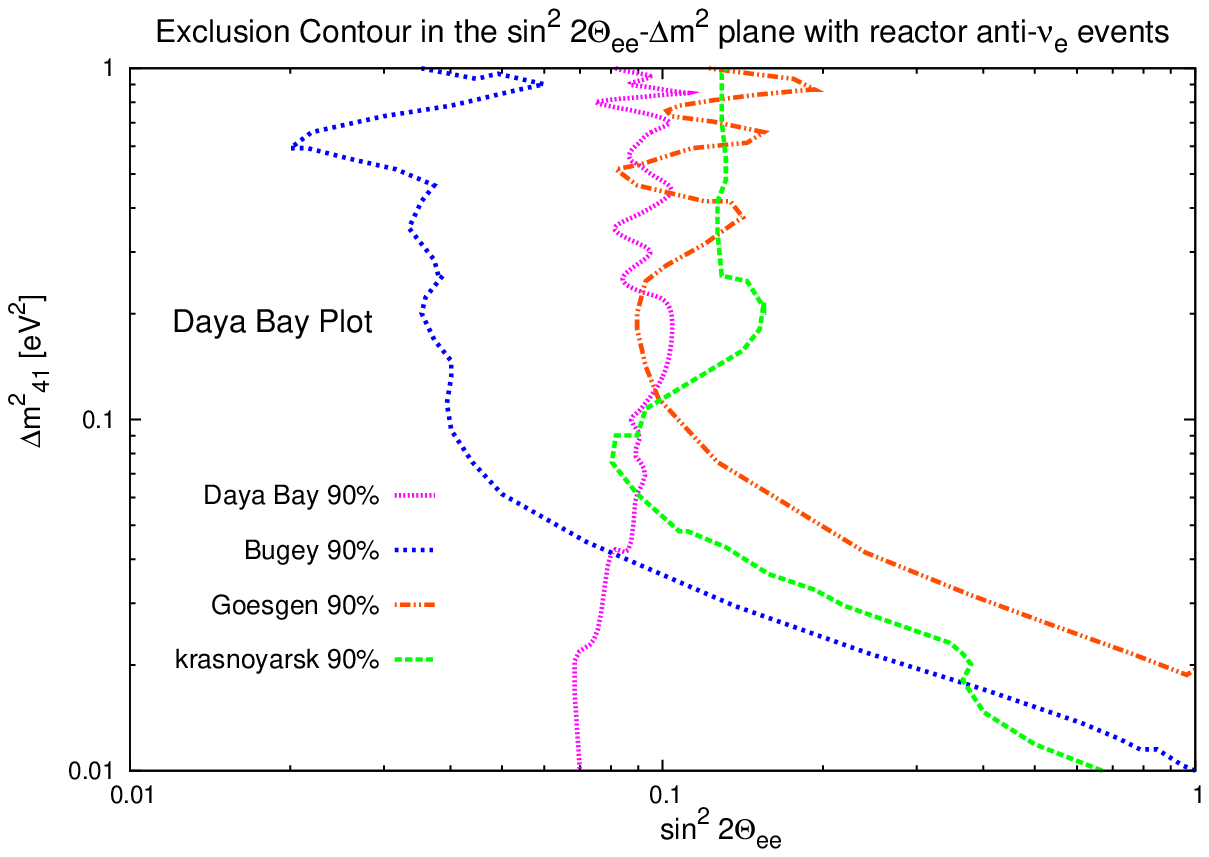}  & \tabularnewline
\end{tabular}
\par\end{centering}

\caption{Comparision of RENO and Daya Bay exclusion plots with other reactor
experiments at 12$\%$ resolution using modified errors. }

\end{figure}

\qquad{}\qquad{}In Fig.2 and 3, we show the comparision
of the exclusion plots for Double Chooz, RENO and daya Bay experiments
with existing reactor experiments using modified errors. Our results
for all the three experiments show better exclusion regions
for sterile neutrino oscillation parameters than Gosgen and Krasnoyarsk in the range 
$\Delta{m}_{41}^{2}$=0.01 to 1 $eV^{2}$. Our bounds are an improvement over Bugey in
the range $\Delta{m}_{41}^{2}$=0.01 to 0.05 $eV^{2}$ but in the $\Delta{m}_{41}^{2}$ region above this, Bugey gives better bounds. From these curves, we see that $\sin^{2}2\theta_{ee}>0.1$ and $\sin^{2}2\theta_{ee}>0.08$
is excluded for sterile oscillation in the $\Delta{m}_{41}^{2}$=0.01
to 1$eV^{2}$ region for Double Chooz and RENO respectively. The Daya Bay
exclusion bound in the region $\Delta{m}_{41}^{2}$=0.1 to 1 $eV^{2}$ is found
to be nearly $\sin^{2}2\theta_{ee}>0.07$.

\section{Conclusions}

From the above results, we can draw the following conclusions:

\begin{itemize}

\item The sensitivity of reactor experiments like Double Chooz, Daya Bay and RENO to sterile oscillation parameters
is significantly dependent on the systematic errors, detector resolutions and uncertainties of each experiment. 
The dependence is especially strong on the correlated reactor-related errors and the normalization uncertainty. 

\item Because of the multiple detectors and baselines in each of these experimental set-ups, it is possible to
have partial cancellations of the experimental errors, especially of the correlated errors, which is beneficial 
in giving us better parameter sensitivities.

\item In an analysis with duly reduced values of errors, it is possible to obtain better bounds with these set-ups than those from many of the older reactor experiments, in spite of the fact that the relatively higher baselines in this case are less suited for a determination of oscillations in the $\Delta m_{14}^2 = 1$ eV$^2$ range. For the latter reason, we find that we obtain better bounds in the low region $0.01 < \Delta m_{14}^2 < 0.05$ eV$^2$.
\end{itemize}

It may be noted that this region lies below the present best-fit range for sterile-scale oscillations. However, the tension between the sterile parameter bounds from different sets of experimental data like the appearance and disappearance experiments indicates that there is still significant uncertainty on the favored region. It is possible that the global fit regions may shift appreciably in the future. In view of this, these results are significant in showing that present reactor experiments like Double Chooz, Daya Bay and RENO may be able to give improved sterile oscillation bounds in specific regions of the parameter space. Such signatures may contribute in modifying the overall picture of sterile oscillation parameter sensitivity. 


\vskip.25in

\textbf{Acknowledgements}

We would like to thank Raj Gandhi for extensive discussions and suggestions. We thank the XI Plan Neutrino Project at HRI, Allahabad, for providing financial assistance to visit HRI, during which major parts of the work have been carried out. KB and PG would like to thank Thomas Schwetz for his useful comments and discussions. KB also thanks UGC SAP program (to Physics Department, Gauhati University) for providing financial assistance.

\end{document}